# Structure and thermoelectric properties of single-phase isocubanite $CuFe_2S_3$ synthesized by mechanical-alloying and ultra-fast microwave radiation technique

Tristan Barbier,*[a] Bhuvanesh Srinivasan,[b,c] David Berthebaud,[a,b] Volker Eyert,[d] Raymond Frésard,[a] Rodolphe Macaigne,[a] Sylvain Marinel,[a] Oleg I. Lebedev,[a] Emmanuel Guilmeau,[a] and Antoine Maignan[a]

[a] *CRISMAT, Normandie Univ., ENSICAEN, UNICAEN, CNRS, 14000 Caen, France*
[b] *CNRS-Saint Gobain-NIMS, IRL 3629, Laboratory for Innovative Key Materials and Structures (LINK), National Institute for Materials Science (NIMS), 1-1 Namiki, Tsukuba 305-0044, Japan*
[c] *WPI International Center for Materials Nanoarchitectonics (WPI-MANA), National Institute for Materials Science (NIMS), 1-1 Namiki, Tsukuba 305-0044, Japan*
[d] *Materials Design SARL, 42, Avenue Verdier, 92120 Montrouge, France*

* **Correspondence**: tristan.barbier@ensicaen.fr

**Abstract**

The current state-of-the-art thermoelectric materials are generally composed of expensive, scarce, and toxic elements. In this respect, copper-based sulfide compounds have emerged as viable alternatives. Herein, we report for the first time the successful synthesis of single-phase cubic isocubanite $CuFe_2S_3$ using mechanical-alloying combined with microwave-assisted synthesis. The isocubanite phase synthesized *via* this ultra-fast out-of-equilibrium process exhibits a maximum thermoelectric figure of merit, $zT_{max} \sim 0.14$ at 673 K. Besides the thermoelectric properties, insights about the structure of isocubanite, based on the refinement of X-ray diffraction data and first principles calculations, are also investigated in detail. It confirms that the Cu-Fe cations in synthetic isocubanite to overwhelmingly occupy the 4$d$ sites of the cubic structure in an inherently disordered fashion.

**Keywords**: Thermoelectrics; Isocubanite; Microwave synthesis; Single-phase; Density of states.

## 1. Introduction

Thermoelectric (TE) power generation can achieve significant energy saving by converting waste heat, generated in many industrial processes, into fruitful electrical energy.[1,2] However, the best performances in low and medium temperature ranges have so far been obtained for thermoelectric metal tellurides/selenides such as PbTe, $Bi_2Te_3$, PbSe, SnSe, SnTe, GeTe, $AgSbTe_2$ compounds, with maximum figure of merit, $zT$, higher than the unity in the temperature range of 350 – 900 K.[3–8] The thermoelectric material's performance is represented by the figure of merit, $zT = \sigma S^2 T/\kappa$, where $\sigma$, $S$, $T$, and $\kappa$ stand for the electrical conductivity, the Seebeck coefficient, the thermal conductivity and the absolute temperature, respectively. The generally complex and expensive synthesis procedures, together with the toxicity, cost and unavailability of those elements, present serious impediments for large-scale commercialization or practical applications. In this context, research efforts are focused on the development of processing methods and thermoelectric materials, that can exhibit high performance and be produced on a larger scale at an economically viable cost.[9,10]

The sulfide minerals are the primary source of world supplies of a wide range of metals. They are the most important ore minerals,[11] thus making them highly relevant for large-scale industrial applications. In addition to their concentrated occurrence in ore deposits and mineralization areas, a limited number of sulfide minerals are found as accessory minerals in rocks. Considering those advantages, some emerging researches on thermoelectrics have focused on copper-based sulfides,[12] which have, for the most part, the advantage of comprising eco-friendly and earth-abundant elements. The obtained zT values in $Cu_{2-x}S$ are quite promising.[13,14] However, from the point of view of device/module fabrication, the electromigration of copper weakens the thermoelectric materials' stability and durability.[15] To overcome this issue, the presence of other metallic elements (M) in the structure, besides copper, can block electromigration. It has opened the route for the exploration of ternary and quaternary copper-based sulfides. Among this large family of compounds, n-type transport behavior is obtained for low Cu content, i.e., Cu/M ratio ≤ 1, as in $Cu_4Sn_7S_{16}$ [16] and chalcopyrite $CuFeS_2$-type sulfides and their derivates $CuFe_{1-x}In_xS_2$ [17] and $Cu_{1-x}Zn_xFeS_2$,[18] where the Fe 3$d$ states play a crucial role in contributing to the electronic band structure of these n-type materials. p-type behavior is found in a broad range of materials with higher Cu content, *i.e.*, Cu/M ratio ≥ 1, as observed in $Cu_2ZnXS_4$ (X = Sn and Ge),[19,20] to copper-rich sulfides (Cu/M ratios between 2 – 5) like $Cu_2SnS_3$[21,22] and tetrahedrites $Cu_{12}Sb_4S_{13}$,[23–26] colusites $Cu_{26}V_2Sn_6S_{32}$,[27–30] bornites $Cu_5FeS_4$,[31–33] and $Cu_5Sn_2S_7$.[34] Recently, ternary Cu-based thiospinel materials with a general composition of $CuM_2S_4$ (M = Ti, Cr, Co, etc.) have also gained some prominence due to their promising TE properties.[35–38]

Among the wide family of copper-based sulfide compounds, cubanite ($CuFe_2S_3$) is a well-known weakly ferromagnetic mineral with an orthorhombic structure.[39] However, when heated above 473 K, the orthorhombic cubanite irreversibly transforms to a face-centered cubic polymorph, termed as isocubanite.[39,40] Analogous to chalcopyrites ($CuFeS_2$), the isocubanite can be portrayed as a tetragonal close-packed stacking of S anions, which occupy the 4$a$ (0, 0, 0) crystallographic site. The cations (Cu, Fe) are known to be randomly distributed over the two structurally equivalent tetrahedral sites 4$c$ (1/4, 1/4, 1/4) and 4$d$ (3/4, 3/4, 3/4) under its mineral occurrence,[39] while its synthetic form only exhibits a random distribution over one of those sites.[41] Due to the aforementioned irreversible transition, no successful attempt to synthesize pure/single-phase cubanite in the orthorhombic form under laboratory conditions has been reported so far using the conventional solid-liquid-vapor reaction in vacuum-sealed tubes (melt processing route). Even in nature, the orthorhombic cubanite generally occurs in the matrix of chalcopyrites ($CuFeS_2$) and pyrrhotites ($Fe_7S_8$).[42] Nevertheless, Chandra *et al*.[42] succeeded in obtaining orthorhombic cubanite without any chalcopyrite phase (albeit with the presence of some secondary pyrrhotite $Fe_7S_8$ phase) using microwave radiation. On the other hand, to the best of our knowledge, not many attempts have been reported so far to synthesize pure cubic cubanite (*i.e.*, isocubanite) compounds. The only available results are from our recent work, where some promising thermoelectric properties ($zT$ ~ 0.13 at 700 K) have been reported in bulk isocubanite containing minor impurities, prepared by traditional melt processing route.[41] To stabilize cubic isocubanite and investigate its TE properties, we have used microwave-assisted synthesis and succeeded in preparing a single-phase compound.

## 2. Materials & Methods

*Synthesis*

Isocubanite single-phase samples were synthesized by a three-step process: firstly, the mechanical-alloying of the starting elements, followed by microwave heating, and finally consolidation by Spark Plasma Sintering (SPS).

Pure powdered starting elements from Alfa Aesar, namely Cu (99.0 %), Fe (99.5%), and S (99.5 %), have been used in a stoichiometric ratio and then placed in a 20 mL tungsten carbide jar containing 5 mm diameter tungsten carbide balls. This assembly (jar, balls, and precursor powders) was then placed in a high-energy ball mill (Fritsch Pulverisette 7 Premium Line). The powders were then processed for two periods of 15 min at a speed of 500 rpm with a time-out of 1 min between each period.

After completing the milling program, the resulting pulverized powder was cold pressed (cylindrical shape) and placed in a glassy carbon crucible (used as susceptor). Crucible and pellet were then placed in a silica tube, sealed under argon atmosphere for preventing any oxygen contamination. Then, this assembly was heated using microwave radiation (2.45 GHz generator - GMP30K, SAIREM, France). The temperature raised to 953 K in few seconds without holding time was measured by an infrared pyrometer (5G-1007 - IRCON, USA). The resulting microwave radiated pellet was then crushed and sieved down to 200 μm before being sintered by Spark Plasma Sintering (HP D 25/1, FCT, Germany) at 673 K for 15 min under a pressure of 63 MPa, thus resulting in a dense disc-shaped pellet with theoretical density > 95%.

*Powder X-ray diffraction*

The crystallographic quality of the densified samples was then determined by using X-ray powder diffraction (XRD). The data were collected using a PANalytical X'Pert Pro diffractometer (θ – 2θ Bragg-Brentano mode) using Co K$_{α1}$ ($\lambda$ = 1.7889 Å) radiation. Indeed, Co-anode based X-ray tube was used instead of Cu-anode for reducing the fluorescence background. Data were collected over the angular range of $10 \leq 2\theta/° \leq 100$ with a step size of 0.01° and a step time of 2 s. Rietveld refinement was performed using the WinPLOTR program included in the FullProf suite software.[43,44]

*Microscopic analysis*

Transmission electron microscopy (TEM), including electron diffraction (ED), high angle annular dark-field scanning TEM (HAADF- STEM), and energy-dispersive X-ray (EDX) spectroscopy analysis, was performed using JEM ARM200F cold FEG double aberration-corrected microscope operated at 200 kV and equipped with ORIUS CCD camera, CENTURIO EDX detector and QUANTUM GIF. TEM specimen was prepared by crushing down the sintered material in an agate mortar with ethanol and depositing the suspension on the Ni holey carbon grid.

*Electrical and thermal transport measurements*

Ulvac Riko ZEM-3 system working under the partial pressure of helium simultaneously measured the temperature dependence of the electrical resistivity and Seebeck coefficient on bar-shaped samples. LFA-457 (Netzsch) was used to measure the thermal diffusivity on square-shaped samples. The total thermal conductivity was obtained by the product of thermal diffusivity, heat capacity (measured with Netzsch DSC system operating under $N_2$ flow), and the measured density (measured using Archimedes' principle). All the transport measurements were performed from 323 K up to 673 K.

*Computational procedures*

All calculations were based on density functional theory (DFT)[45,46] with exchange-correlation effects accounted for by the generalized gradient approximation (GGA).[47] Two complementary tools were employed: In a first step, the Vienna ab-initio simulation program (VASP) as implemented in the MedeA® computational environment of Materials Design was used to perform total-energy and force calculations aiming at an optimization of the structures.[48–50] As an exchange-correlation functional the PBEsol scheme was selected.[51] The single-particle equations were solved using the projector augmented wave (PAW) method[52,53] with a plane-wave basis with a cutoff of 384.080 eV. The Brillouin zone was sampled using a Monkhorst-Pack mesh with $7 \times 7 \times 3$ $k$-points.[54] Once the optimized structures were known, calculations of the electronic structure were carried out using the full-potential augmented spherical wave (ASW) method in its scalar-relativistic implementation.[55,56] Exchange and correlation effects were considered using the GGA parametrization of Wu and Cohen.[57] In the ASW method, the wave function is expanded in atom-centered augmented spherical waves, which are Hankel functions and numerical solutions of Schrödinger's equation, respectively, outside and inside the so-called augmentation spheres. Additional augmented spherical waves were placed at carefully selected interstitial sites to optimize the basis set and enhance the variational freedom. The choice of these sites and the augmentation radii were automatically determined using the sphere geometry optimization algorithm.[58] Self-consistency was achieved by a highly efficient algorithm for convergence acceleration[59] until the variation of the atomic charges was smaller than $10^{-8}$ electrons and the variation of the total energy was smaller than $10^{-8}$ Ryd. Brillouin zone integrations were performed using the linear tetrahedron method[60] with $8 \times 8 \times 3$ $k$-points. Finally, the thermopower was calculated within Boltzmann theory in the relaxation-time approximation. As usual, the relaxation time was assumed to be independent of the $k$-point and band index, in which case it cancels from the Seebeck coefficient. Calculation of the transport properties is fully integrated into the ASW method. Details of the calculational approach and its implementation are given in the previous reports.[56,61]

## 3. Results & discussion

### 3.1 Structural properties

XRD patterns of as-milled powder is displayed in Figure 1 (Top). Main diffraction peaks of Fe and S can be observed, together with broad and less intense diffraction peaks of CuS. The crystallization of CuS after only 30 min of mechanical-alloying is expected thanks to its low formation energy.[62] Similar observations have been recently reported in $Cu_{2-x}S$, $Cu_2SnS_3$, tetrahedrite and colusite $Cu_{26}V_2Sn_6S_{32}$.[63–66]

Simulated XRD patterns of orthorhombic and cubic cubanites ($CuFe_2S_3$) along with that of chalcopyrite ($CuFeS_2$) patterns are depicted in Figure 1 (labeled as modelization). Experimental powder XRD patterns of isocubanite compounds synthesized by the conventional solid-liquid-vapor reaction in vacuum-sealed tubes (extracted from our previous work[41]) and by microwave radiation (from this work) are displayed in Figure 1 (labeled as exp. data) for comparison. XRD patterns show that both compounds crystallize within the cubic polymorph with space group $F4\text{-}3m$ (n° 216). However, the melt-processed isocubanite pattern shows some additional reflections at 57.5° and 69° (pointed out with # marks) corresponding to that of the chalcopyrite phase. Such intimately intergrown chalcopyrite secondary phase (along with some traces of pyrrhotites) in cubanites seems to be common, as previously reported in the literature.[42,67] In the case of microwave processed sample, no extra peaks associated to secondary phases are observed, and all the peaks can be indexed in the cubic polymorph. Hence, it can be stated that, using mechanical alloying followed by microwave heating, a pure isocubanite phase is successfully synthesized for the first time. The nucleation of CuS during mechanical alloying, combined with the small grain/crystallite size, favors the reactivity and fast crystallization during microwave synthesis to produce highly pure cubic cubanite in few minutes.

In the past, domestic microwave ovens have been used in an attempt to synthesize pure orthorhombic cubanites; however, it still resulted in other minor secondary phases, even when processed for just a few seconds with a power of 600 – 900 W.[42,67] In this work, the single-mode radiation together with the carbon glass susceptor used at a capacity of 250 W might have helped to the formation of the single-phase isocubanite. Indeed, those microwave mono-mode radiations allow obtaining super-fast heating and cooling rates inside the crucible, *i.e.*, less than 1 min for reaching 953 K from room temperature. It is then naturally cooled down to room temperature, (*i.e.* in less than 5 min), leading to a quenching allowing non-equilibrium thermodynamic conditions to be reached. Out-of-equilibrium processes such as flash (hybrid flash-SPS) processing,[68–71] which employs thermal runaway to achieve ultra-fast heating and cooling rates, are known to produce single-phase compounds by exceeding the equilibrium solubility (*i.e.* solubility under general equilibrium conditions).

The Rietveld refinement of XRD pattern of the sintered sample prepared from the mechanical alloying/microwave radiation powder, depicted in Figure 2, confirms a high purity and crystallinity sample. The refined parameters are given in Table 1. As two structurally equivalent tetrahedral sites, termed 4*c* (1/4; 1/4; 1/4) and 4*d* (3/4; 3/4; 3/4) can be defined for both copper and iron atoms in the isocubanite phase, refinements were thus performed using both crystallographic sites.[41] However, the XRD technique cannot predict the obtained crystallographic configuration of this isocubanite phase due to the close electronic density of Cu and Fe atoms. Refinements using the 4*c* site for both Cu and Fe atoms and refinements using the 4*d* one has led to very similar crystallographic parameters, as it can be observed from the values summarized in Table 1. Irrespective of the Cu and Fe crystallographic sites being used, similar crystallographic parameters have been obtained ($a$ = 5.2994 Å, close to those previously reported in the literature).[39,41,67] Indeed, all metal atoms can either be found on 4*c* or 4*d* (both situations are equivalent, as they are related to each other by an inversion).[39] To assess the random distribution of metal atoms within the 4*d* tetrahedral site, electronic band calculations of isocubanite compound have been undertaken (next section).

HAADF-STEM study and STEM-EDX mapping were performed to investigate the structural and chemical homogeneity at a more local scale. Electron diffraction patterns (depicted in Figure 3a) highlight well-defined spots, which are characteristic of a highly crystallized compound, and can be fully indexed based on a cubic $F\bar{4}3m$ structure with a cell parameter of 5.2994 Å (as determined by the XRD analysis). STEM-EDX mapping (Figure 3b) evidences a perfect chemical homogeneous distribution of all elements with a composition close to the nominal one. No segregation of any element was observed. High-resolution HAADF-STEM imaging (Figure 3c) confirms the high crystallinity and the absence of any point defects and/or deformations. However, some twinning defects, which are typical [011] twinning for a cubic structure with <111> twinning plane, have been observed (Figure 3d).

*3.2 Electronic structure & density of states*

To capture the random distribution of copper and iron within the cation lattice of the zinc-blende structure, a supercell consisting of three cubic cells stapled on top of each other was built; they are the smallest possible supercells allowing for the correct stoichiometry of the compound. The resulting twelve cation sites were occupied by four Cu and eight Fe atoms, all found in 4*d* position of the underlying cubic cell. Out of all possible atom arrangements, three candidates with different degrees of clustering of the Cu atoms were selected. These three different structures are displayed in Figure 4. All the structures were fully optimized, *i.e.*, the lattice vectors and the internal coordinates were relaxed using VASP. The calculated lattice parameters, cell angles, total energies of structures, and realization probabilities are summarized in Table 2. Obviously, the three structures contain different degrees of clustering of the Cu atoms, which are also reflected by the calculated pair correlation functions. While structure 3 is close to a homogeneous distribution of Cu, structures 1 and 2 reveal an overall increased number of Cu-Cu pairs within the cation sub-lattice. The structural differences can be attributed to an increased bonding within Cu-Cu pairs compared to all other cation combinations. This increased bonding causes the deviation of the angle $\gamma$ in structure 1, which is the only structure with two Cu atoms in an *ab*-plane. Also, we observe a more substantial decrease of the *c*-axis in structures 1 and 2, which have more Cu-Cu pairs in neighboring *ab*-planes than structure 3 with its more homogeneous distribution of the Cu sites. Specifically, the change in c lattice parameter is almost proportional to the number of Cu-Cu bonds parallel to the c-axis.

This preference of Cu-Cu pairs in the cation sub-lattice is reflected by the calculated total energies of the optimized structures. Nevertheless, let us emphasize that the three calculated configurations are very close in energy. It is therefore of interest to determine their statistical distribution at temperatures close to the actual synthesis temperature (953 K) or above, taking 1500 K, for instance. As presented in Table 2, the lowest in energy configuration may indeed be found with the largest probability (0.695), but the probability of finding any of the other two is not negligibly small, and it increases with increasing synthesis temperature. Hence, the actual structure of the sample is a mixture of all three structures.

To address the electronic structure, calculations of the (partial) densities of states were performed using the ASW method. The results are shown in Figure 5. As expected, the partial densities of states of the three structures are very similar. While the lower valence band in the energy range from -7 to -4 eV is dominated by the S 3*p* states, Fe and Cu 3*d* states are mainly found between -3 and +3 eV. In particular, the Cu 3*d* states show a sharp peak at about -3 eV due to the non-bonding $e_g$ states and about 1 eV broad (approx. from -3 eV to -2 eV) $t_{2g}$ bands above. These bands mediate the strong $\sigma$-like bonding with the S 3*p* orbitals. This overall structure is similar to the Fe 3*d* states. Their $e_g$ manifold is narrow and found between -2 and -0.5 eV, whereas the much broader $t_{2g}$ bands extend from -2 to +3 eV and come with a considerably larger S 3*p* admixture. From all calculations, we obtain rather similar orbital occupations, which are Cu 4*s* 0.44, 4*p* 0.43, 3*d* 9.18; Fe *4*s 0.35, 4*p* 0.39, 3*d* 6.14; and S 3*s* 1.55, 3*p* 3.17. Furthermore, the cations were found to carry no magnetic moment.

The respective computed Seebeck coefficients for the three structures that are shown in Figure 4 are displayed in Figure 6, where the three different Cartesian components (*xx*, *yy*, *zz*) are distinguished. In addition, the averages of these components are shown (solid red line). The results reveal striking differences in the three structures due to the different arrangements of the Cu atoms.

### 3.3 Thermoelectric properties – *comparison of melt-processed vs. microwave synthesized isocubanite*

The electrical and thermal transport properties of melt-processed and microwave synthesized isocubanite and sintered by SPS are shown in Figure 7. It must be again noted that the microwave route yielded pure / single-phase isocubanite, while the melt-processing route resulted in some additional traces of chalcopyrite and pyrrhotite. Though not consistent throughout the whole temperature range, the electrical conductivity stays at the same level or slightly decreases with temperature until 473 K, and then slightly increases, and these two compounds (with cubic structure) exhibit $\sigma$ higher than 6 × $10^4$ S/m. This contrasts with the low electrical conductivity in the orthorhombic cubanites,[72] which makes those orthorhombic structures thermoelectrically less attractive. Considering that the chalcopyrites and pyrrhotites are an order of magnitude lower in $\sigma$ when compared to isocubanite,[73,74] one would expect the melt-processed sample with those additional phases to exhibit lower $\sigma$ than the microwave synthesized pure isocubanite. However, that is not the case here, and at this juncture, it would be difficult to pinpoint the exact reason for this anomaly, as the role of the contribution of individual secondary phases to the overall electrical transport is not the only determining factor; first, a slight difference in the exact stoichiometry between the samples can be responsible for a slight change in the charge carrier concentration. In that respect, the higher resistivity of the present sample is consistent with a lower carrier concentration which is also consistent to its larger S absolute value (see below). Of course, other factors such as electron diffusion, scattering arising due to ionized impurities and/or electron-phonon interactions, and the correlation of elemental bond distributions, all of which can accordingly influence the charge carrier mobility (and ultimately $\sigma$). The electrons being the dominant charge carriers, the Seebeck coefficient is negative for these compounds (Figure 7b). This experimentally observed *n*-type transport is consistent with the DFT computations, *i.e.*, the Fermi-level ($E_F$) position in the DOS calculations (Figure 5). The thermopower of these isocubanite exhibit a weak temperature dependence throughout the measured temperature range (323 – 673 K), roughly creeping around -65 to -70 µV/K for melt processed isocubanite, and -77 to -80 µV/K for the microwave processed isocubanite. The comparison of the experimental Seebeck coefficient (Figure 7b) to the ones that were simulated (Figure 6), seems not to disprove configuration 2 (Figure 4) to be predominantly realized, as, regarding S_zz as the component of the thermopower that is best captured by our calculations on this supercell, it yields the value closest to experiment. The weak temperature dependence of the electrical transport properties hints at a strong disorder in the 4*c* or 4*d* sites of the isocubanite. The power factor ($PF = \sigma S^2$) for the compounds synthesized by both techniques is similar (Figure 7c), but the microwaved sample slightly outperforms the melt-processed one at the lower temperature ranges. The *PF* increases with temperature and reaches a maximum of $5 \times 10^{-4}$ W/mK$^2$ at 673 K, which can be considered reasonably decent for sulfide-based materials. The thermal conductivity, which continuously decreases with temperature, is marginally lower for the microwaved sample than the melt-processed one (Figure 7d). This difference could be due to their respective differences in their electronic contribution ($\kappa_e$) to the thermal conductivity. But the lower thermal transport in isocubanite, in general, could also be due to their lower lattice contribution ($\kappa_{latt}$) arising not only from the lattice twinning (interfaces, as shown in Figure 3d) factor but also from the local disorder due to the random distribution of Cu and Fe cations in the structure. The fast heating and cooling rates employed during the micro-wave processing can suppress the grain growth and enhance the boundary scattering of heat-carrying phonons in the inter-grain region, thus resulting in lower thermal transport. The figure of merit, which almost linearly increases with temperature (Figure 7e), reaches a maximum of ~0.14 at 673 K for the microwave radiated isocubanite. It is comparable or a touch higher than its melt-processed counterpart.

To put it in a nutshell, given their comparable electrical and thermal transport properties, there is not much noticeable/significant variation in the overall thermoelectric properties of pure (single-phase) microwaved isocubanite when compared to the multi-phased melt-processed compound. It is remarkable that the intrinsic thermal conductivity found for the pure isocubanite is low, reaching only 2.7 W/mK at 525K, and thus even lower than that of the "isocubanite" sample prepared by melt processing. This demonstrates that for the latter, the inclusions of impurity phases (chalcopyrite and pyrrhotite) are responsible for only a minor contribution to explain the low value of the "isocubanite" thermal conductivity reported in ref. [37]. Accordingly, the simple, fast, and effective microwave radiation method has enabled the production of pure isocubanite without any compromise on the thermoelectric performance when compared to the laborious and time-consuming classical vacuum sealed-tube melt-processing method. Hence, moving forward, mechanical-alloying combined with microwave-assisted synthesis appears as quite a useful method to stabilize out-of-equilibrium phases.

## 4. Conclusion

A novel and faster synthesis method, which combines ball milling, microwave radiation, and SPS sintering, has been successfully adopted to produce dense $CuFe_2S_3$ isocubanite. The structural analysis confirmed the high purity (single-phase) of the produced isocubanite and evidenced a perfect chemical homogeneous distribution of all elements with a composition close to the nominal one. For this cubic polymorph, the structural refinements did not evidence ordering of the metal atoms (Cu, Fe) into $4d$ tetrahedral crystallographic site. From the several theoretical structural models computed (DFT) based on different degrees of clustering of the Cu atoms, the structure with a homogeneous distribution of Cu-Cu pairs seems to be more stable and coherent with the experimentally obtained thermopower. As the other two considered structures are very close in energy, it is most likely that all three of them are statistically distributed at the actual synthesis temperature. This peculiarity of the isocubanite results in a Cu and Fe cation distribution on the $4d$ sites that is dominated by a homogeneous distribution of Cu-Cu pairs of atoms. From the calculations of their electronic structure, isocubanite are $n$-type conducting and paramagnetic. The simpler and faster microwave route yielded pure isocubanite with comparable (or marginally higher) thermoelectric performance ($zT_{max}$ ~ 0.14 at 673 K) as that of multi-phased isocubanite synthesized by the traditional vacuum sealed-tube melt-processing method.

## Conflicts of interest

There are no conflicts to declare.

## Acknowledgment

B.S acknowledges the Japan Society for the Promotion of Science (JSPS) for the postdoctoral fellowship (P19720). B.S and D.B appreciate Prof. Takao Mori (NIMS) and Dr. Jean-François Halet (CNRS) for their constant support. The authors would also like to extend their gratitude to Dr. Julien Varignon (CRISMAT) for his inputs.

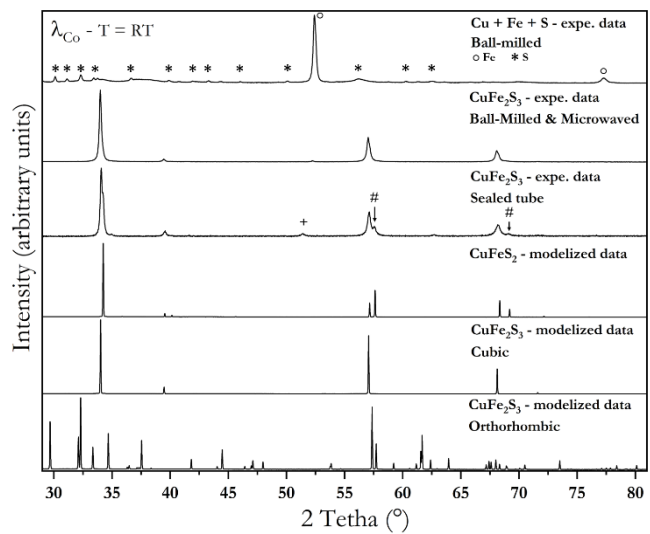

**Figure 1**. Powder X-ray diffraction patterns showing (from bottom to top) simulated orthorhombic and cubic $CuFe_2S_3$, and $CuFeS_2$ pattens (modelized based on their ICSD cards); $CuFe_2S_3$ synthesized using sealed-tube and microwave processes, and the elemental precursors after ball-milled process. * points out at the peak arising due to the sample holder, # points out at the peak shoulder corresponding to the secondary chalcopyrite ($CuFeS_2$) phase.

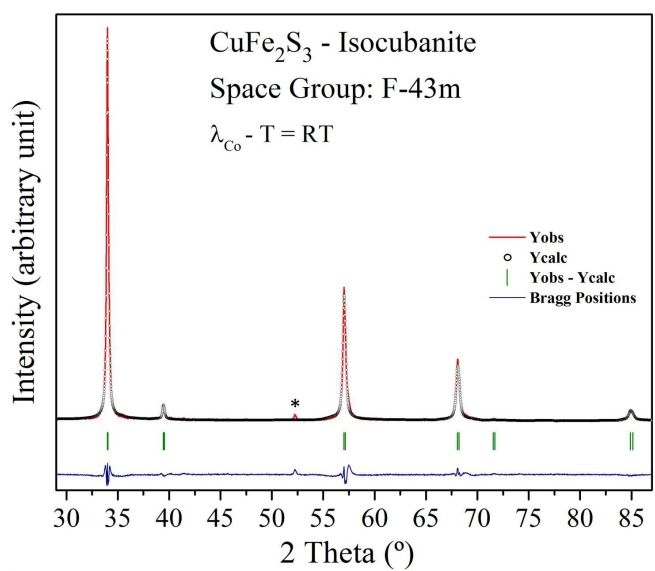

**Figure 2**. Rietveld refined powder XRD pattern of $CuFe_2S_3$ isocubanite synthesized *via* microwave radiation method (pattern collected on SPS sintered sample). * indicates the peak arising due to the sample holder.

**Table 1**. Rietveld refinement parameters of CuFe$_2$S$_3$ isocubanite synthesized *via* microwave radiation technique and consolidation by SPS. Crystallographic parameters obtained through both refinements, using 4*d* and 4*c* crystallographic sites, respectively, are tabulated (Occupancies for Cu and Fe atoms were fixed to 0.333 and 0.666, respectively).

| Atoms | Site | *x* | *y* | *z* | Biso |
|---|---|---|---|---|---|
| Cu | 4*d* | 0.75 | 0.75 | 0.75 | 1.903(6) |
| Fe | 4*d* | 0.75 | 0.75 | 0.75 | 1.903(6) |
| S | 4*a* | 0 | 0 | 0 | 3.036(6) |
| Reliability factors | \multicolumn{5}{c}{$R_{Bragg}$ = 3.13; $R_F$ = 5.03} |
| Cell parameter | \multicolumn{5}{c}{*a* = 5.2994(7) Å} |
| Cu | 4*c* | 0.25 | 0.25 | 0.25 | 1.903(2) |
| Fe | 4*c* | 0.25 | 0.25 | 0.25 | 1.903(2) |
| S | 4*a* | 0 | 0 | 0 | 3.037(7) |
| Reliability factors | \multicolumn{5}{c}{$R_{Bragg}$ = 3.08; $R_F$ = 4.25} |
| Cell parameter | \multicolumn{5}{c}{*a* = 5.2994(4) Å} |

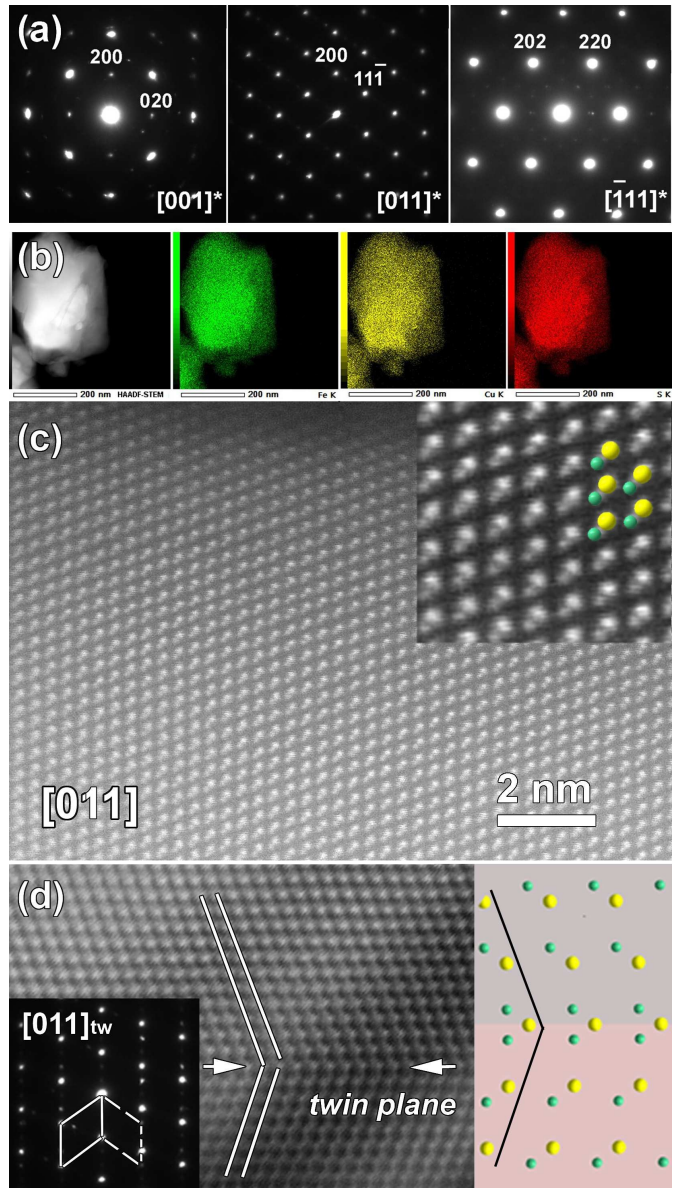

**Figure 3**. (a) Electron diffraction patterns along the three main zone axes of the cubic CuFe$_2$S$_3$ structure; (b) STEM-EDX elemental mapping (Fe, Cu, and S) of selected crystallites; (c) high-resolution [011] HAADF-STEM image. Enlargement of the isocubanite crystal structure containing an overlaid structure model (Fe/Cu: green, S: yellow) is shown as an inset; (d) enlargement of high resolution [011] HAADF-STEM image exhibiting twinning interfaces and schematic representation of the latter interfaces.

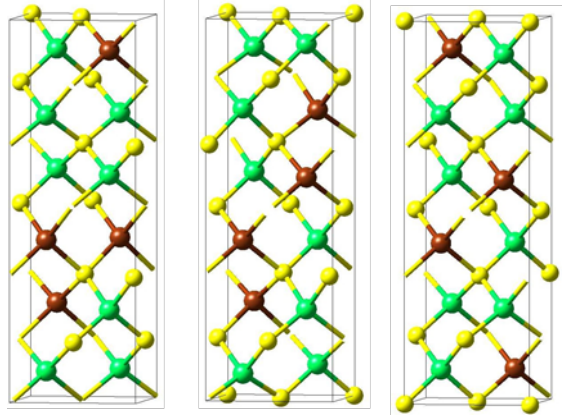

**Figure 4**. Three structures of CuFe$_2$S$_3$, which are distinguished by their occupation of the cation sites by Cu and Fe. Here, the fully optimized structures are shown. For further reference, the structures (configurations) are labeled as 1, 2, 3 from left to right. While configuration 3 is close to a homogeneous distribution of Cu, configurations 1 and 2 reveal the overall increased number of Cu-Cu pairs within the cation sub-lattice. Color code for atoms: S – yellow, Fe – green, Cu – red.

**Table 2**. Calculated lattice parameters, angle $\gamma$, and total energies of the three structures of CuFe$_2$S$_3$ (shown in Figure 4), and their realization probabilities. The angles $\alpha$ and $\beta$ are 90°.

| Configuration | $a$ (Å) | $b$ (Å) | $c/3$ (Å) | $\gamma$ | $E_{final}$ (eV) | Realization probability at 953 K (1500 K) |
|---|---|---|---|---|---|---|
| 1 | 4.9657 | 4.9657 | 4.9796 | 90.44 | −149.1576 | 0.192 (0.251) |
| 2 | 4.9719 | 4.9733 | 4.9470 | 90.00 | −149.2642 | 0.695 (0.569) |
| 3 | 4.9603 | 4.9510 | 5.0157 | 90.00 | −149.1131 | 0.113 (0.180) |

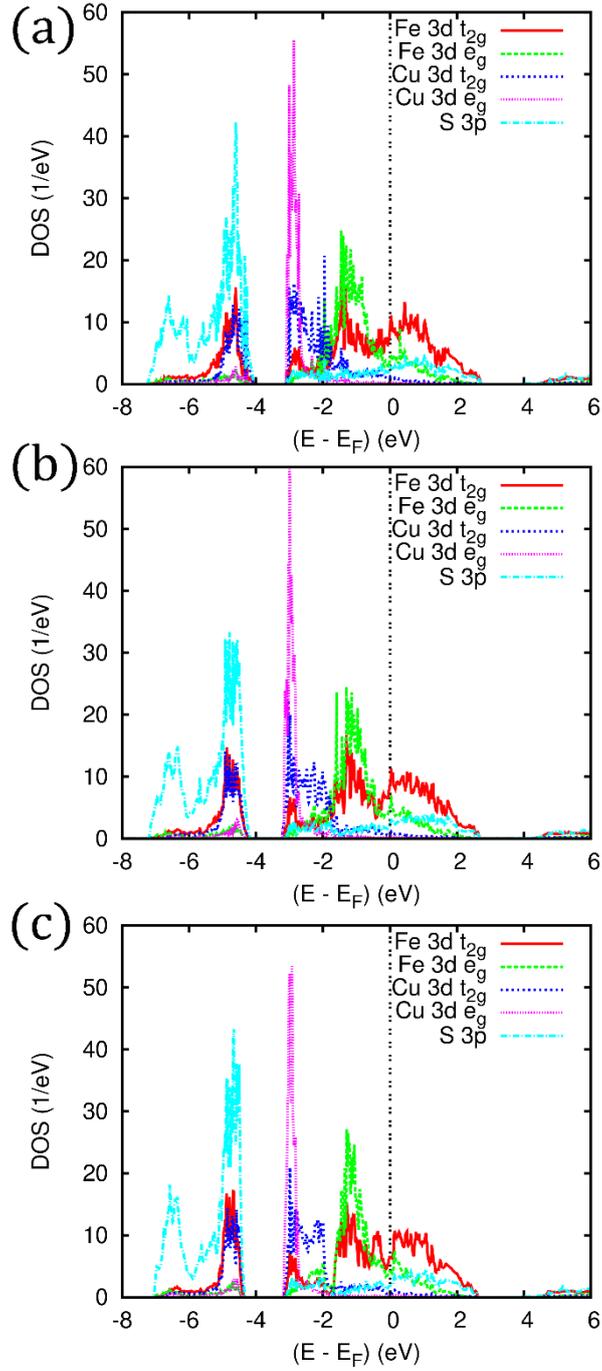

**Figure 5**. Partial densities of states (DOS) for the three different configurations of $CuFe_2S_3$ shown in Figure 4.

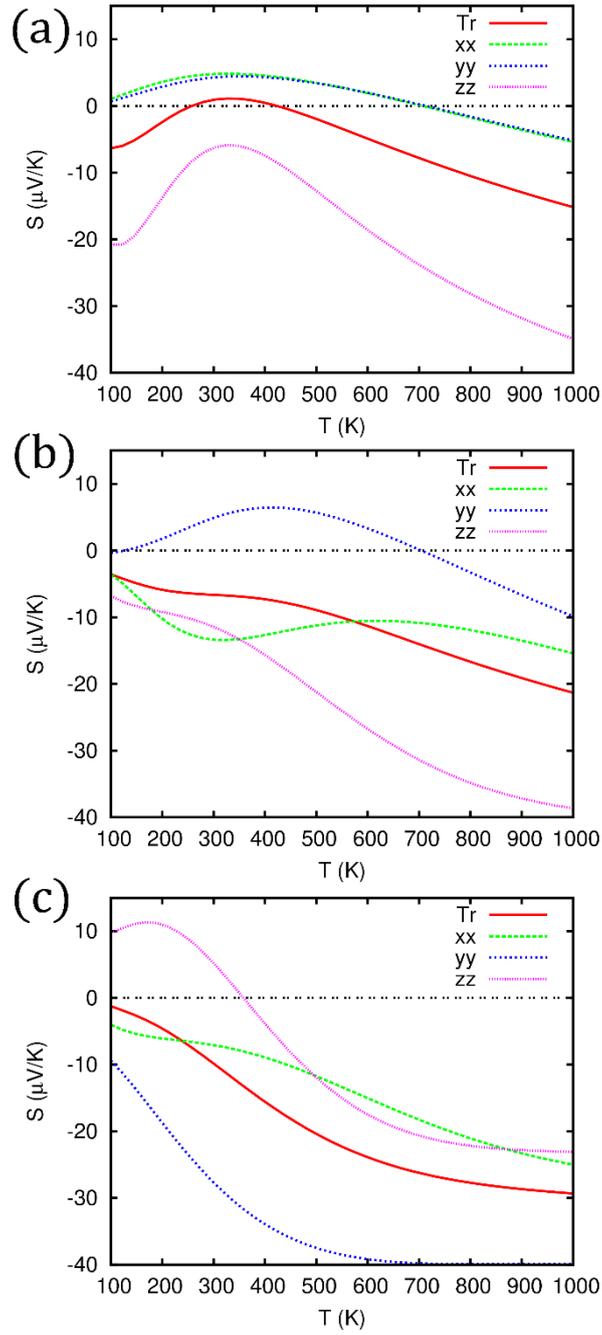

**Figure 6**. Theoretically computed Seebeck coefficient for the three different configurations of $CuFe_2S_3$ shown in Figure 4.

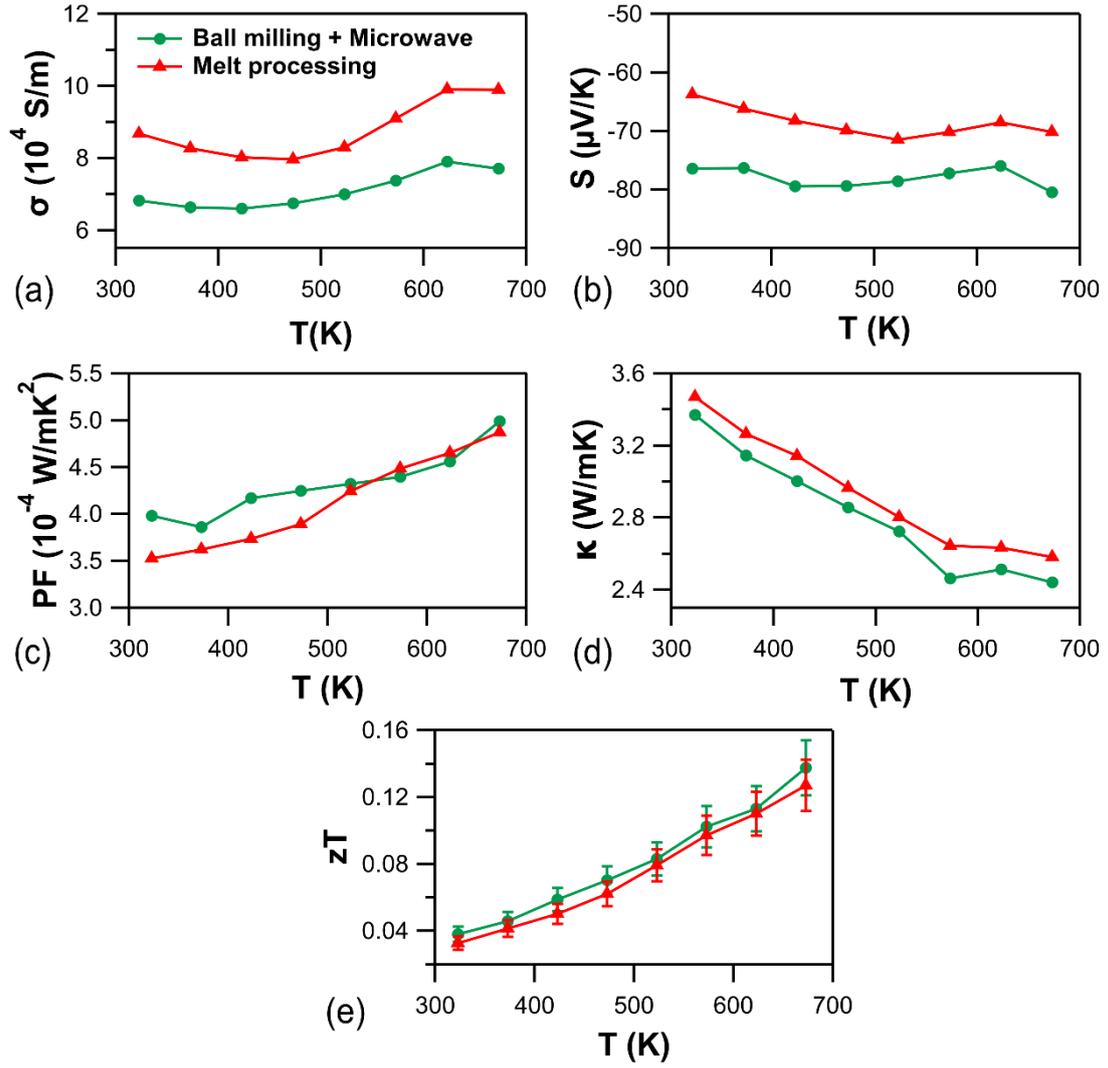

**Figure 7**. Comparison of temperature-dependent transport properties of microwave radiated and melt-processed $CuFe_2S_3$ isocubanite – (a) electrical conductivity, $\sigma$, (b) Seebeck coefficient, $S$, (c) power factor, $PF$, (d) thermal conductivity, $\kappa$, and (e) figure of merit, $zT$.